\documentclass[journal]{IEEEtran}

\IEEEoverridecommandlockouts                              

\usepackage{algorithmic}
\usepackage{amsmath,amssymb,amsfonts}       
\usepackage{amsthm}
\usepackage{array}
\usepackage[english]{babel}                         
\usepackage{empheq}
\usepackage{footmisc}
\usepackage{graphicx}
\usepackage{import}
\usepackage{mathabx}
\usepackage{physics}
\usepackage{textcomp}
\usepackage{transparent}
\usepackage{stackrel}
\usepackage{xcolor}
\usepackage{xfrac,enumitem}
\usepackage{mathrsfs}
\usepackage{flushend}
\usepackage{soul}              

\usepackage{etoolbox}

\makeatletter
\let\IEEEorigbibitem\bibitem

\renewcommand{\bibitem}[1]{%
  \color{black}%
  \IEEEorigbibitem{#1}%
  \ifstrequal{#1}{fusco2014hierarchical}{\color{black}}{}%
  \ifstrequal{#1}{Anderson2025igde}{\color{black}}{}%
  \ifstrequal{#1}{Xu2015zbf}{\color{black}}{}%
  \ifstrequal{#1}{Merigaud2018horizon}{\color{black}}{}%
  \ifstrequal{#1}{fornaro2025filteringST}{\color{black}}{}%
  \ifstrequal{#1}{Fornaro2026host}{\color{myred}}{}%
  \ifstrequal{#1}{MendozaAvila2023parasitic}{\color{myred}}{}%
  \ifstrequal{#1}{Mosquera2023gst}{\color{myred}}{}%
}
\makeatother

\newtheorem{Remark}{Remark}

\usepackage[left=54pt, right=54pt, top=54pt, bottom=54pt]{geometry}

\def\BibTeX{{\rm B\kern-.05em{\sc i\kern-.025em b}\kern-.08em
    T\kern-.1667em\lower.7ex\hbox{E}\kern-.125emX}}
\definecolor{myblue}{rgb}{0.3,0.3,1}
\definecolor{bluetheme}{rgb}{0.17,0.26,0.26}
\definecolor{teblue}{rgb}{0.27,0.36,0.36}
\definecolor{mygreen}{rgb}{0.3,0.7,0.5}
\definecolor{myred}{rgb}{0,0,0}

\newcommand{\pedro}[1]{{\textcolor{myred}{#1}}}
\newcommand{\transp}{^\intercal}
\newcommand{\Rset}{\mathbb{R}}

\newcommand{\mC}{\mathcal{D}}
\newcommand{\mb}[1]{\mathbf{#1}}
\newcommand{\ms}[1]{\boldsymbol{#1}}  

\allowdisplaybreaks    
\title{Symphony: Simple Phase Control for \\ Wave Energy Systems}

\author{Pedro~O.~Fornaro,~Eugenio~M.~Gelos,~Demian~Garcia-Violini,~and~John~V.~Ringwood,~\IEEEmembership{Fellow.}
\thanks{Pedro O. Fornaro is with  Centre for Ocean Energy Research (COER), National University of Ireland, Maynooth.}
\thanks{Eugenio M. Gelos, is with Centre for Ocean Energy Research (COER), National University of Ireland, Maynooth.}%
\thanks{Demián García-Violini is with Consejo Nacional de Investigaciones Cient\'ificas y T\'ecnicas (CONICET), with Departamento de Ciencia y tecnolog\'ia, Universidad Nacional de Quilmes, Argentina, and with the Centre for Ocean Energy Research (COER), Maynooth University, Ireland.}%
\thanks{John. V. Ringwood is with Centre for Ocean Energy Research, National University of Ireland, Maynooth.}
}
\begin{document}
\maketitle

\begin{abstract}
Despite significant advancements in controlling wave energy converters, efficiently achieving real-time energy-maximising control remains a challenge. This paper addresses this obstacle by introducing Symphony, a novel non-optimisation-based (NOB) controller inspired by phase control techniques for wave energy devices. Symphony is a suboptimal control that inherits all the advantages of NOB controllers, such as simplicity of design, implementation, tuning, and use, while additionally providing hard-motion-constrained \pedro{reference trajectories} and close-to-optimal performance across the evaluated sea states, even in the presence of measurement noise and model uncertainty. Symphony is designed using a hierarchical structure: By relying on an implicit Gaussian differential equation that requires an excitation force estimate, an upper loop provides a close-to-optimal motion-constrained velocity trajectory and a feedforward control term. Then, a user-flexible lower loop tracks the desired velocity trajectory by combining feedback and feedforward terms. In this paper, Symphony performance is assessed in terms of power absorption. Statistically representative results are obtained using multiple sea-state realisations across varying peak wave periods, considering measurement noise and radiation model uncertainty. The results presented in this paper demonstrate the \textit{potential} of Symphony, which achieves, in most of the evaluated scenarios, more than 90\% of the power absorption obtained by a spectral (optimisation-based) control solution.
\end{abstract}

\begin{IEEEkeywords}
Wave energy, causal control, optimisation, energy maximisation, simplicity.
\end{IEEEkeywords}
\IEEEpeerreviewmaketitle

\section{Introduction}\label{s:1}
Wave energy has emerged as a promising renewable resource, with a global annual potential estimated between 1 and 10 TWh \cite{Reguero2015resource}. Although the exact figure for extractable wave power remains debated \cite{mork2010assessing}, it is agreed that wave energy systems may play a critical role in reducing carbon emissions by contributing to the energy mix \cite{Vazquez2024}. For instance, by reducing production variability, the integration of wave energy with other renewables could enhance energy system stability, {resilience,} and grid penetration. Despite its promise, high capital and operational costs associated with deployment in the harsh ocean environment hinder the large-scale commercialisation of wave energy converters (WEC) \cite{ringwood2023empowering}.

In this context, control strategies represent a crucial element towards the commercialisation of WECs \cite{Ringwood2020sensitivity}\cite{ringwood2014energymax}. This is because control algorithms are essential to maximising energy capture, while managing operational constraints to protect the device from damage \cite{faedo2017optimal}. Common optimisation-based (OB) controllers, such as model predictive control, spectral methods \cite{Garcia2019spectral}, and moment-based algorithms \cite{faedo2017optimal}\cite{Faedo2021moments}, provide, by resorting to numerical routines, constrained energy-maximising solutions. However, OB strategies are heavily dependent on the system model and, hence, susceptible to variations and uncertainties in the WEC dynamics, which appear during real-time operation. Additionally, in order to appropriately handle constraints, OB controllers require prediction of future wave elevation or excitation force, resulting in computationally expensive routines, and hampering real-time implementation of these techniques. {Importantly,} although reducing the computational complexity of OB algorithms is possible, this may result in solutions that do not satisfy the hard constraint requirements. 

In contrast, non-optimisation-based (NOB) control strategies do not rely on numerical optimisation routines, and provide real-time (suboptimal) implementable solutions. However, none of the existing {NOB} controllers are capable of effectively handling hard constraints, typically relying on controller detuning to retain safe operation \cite{garcia2020simple}. {In addition,} most of the existing {NOB} controllers are dependent on a full WEC model, such as the LiTe-Con \cite{Garcia2020litecon}, the LiTe-Con+ \cite{garcia2023broadband}, or the feedback resonating controller \cite{Bacelli2020complex}. Other {NOB} controllers, designed using an approximate velocity tracking structure (such as the Simple \& Effective controller \cite{fusco2012simple}), are dependent only on the WEC radiation dynamics \cite{Ringwood2020sensitivity}. Overall, in spite of the appealing computational advantages, in terms of energy absorption, {NOB} controllers have not yet reached the effectiveness of OB strategies.

To close the gap in performance between {NOB} and OB control strategies, this study introduces a novel NOB simple-phase-controller (Symphony) that, (i) is conceptually simple, (ii) to the best of the authors' knowledge, is the first NOB controller capable of effectively and robustly \pedro{providing constrained} position and velocity \pedro{references}, and, (iii) Symphony obtains the highest levels of mechanical power absorption among the class of {NOB} controllers studied in~\cite{garcia2020simple}. 

Symphony operates using a hierarchical structure. An upper loop generates a feedforward control force and a suboptimal velocity reference that is proportional to a wave excitation force estimate, effectively phase-matching the WEC velocity with the wave excitation force estimate. Then, a lower loop implements the generated velocity reference by resorting to a combination of feedforward and feedback terms. 

The upper loop provides a motion-constrained velocity reference and a feedforward control term. Forward-invariant trajectories are obtained by solving an implicit Gaussian differential equation (IGDE)~\cite{Anderson2025igde}, meaning that IGDE solutions remain inside a forward invariant set (namely a \textit{safe set}, according to the definition in~\cite{Xu2015zbf}), complying with hard motion constraints (i.e., constraints for both position and velocity), even in the presence of measurement noise and model uncertainty. Additionally, unlike in OB controllers, motion-constrained references are obtained without resorting to forecasting, since only a current wave excitation force estimate is required. While multiple alternatives to estimate the wave excitation force are available in the literature (see, e.g.,~\cite{pena2019critical} and references therein), to ameliorate the effects of measurement noise in the upper loop, the estimator from~\cite{fornaro2024homogeneous} is used and recommended.

Then, in Symphony, the lower loop combines feedforward and feedback terms, as in~\cite{Merigaud2018horizon}\cite{Faedo2024sm2}. The feedforward term provides an average control force required to drive the tracking error close to the origin; then, ideally, the tracking control further reduces the tracking error to a small neighbourhood around zero, ultimately depending on the robustness of the employed tracking controller. In this paper, as a tracking controller, the filtering sliding-mode (SM) control from~\cite{fornaro2025filteringST} is used, owing to its robustness to unmodelled disturbances and measurement noise. Notably, in Symphony, while SM-based approaches~\cite{fornaro2024homogeneous}\cite{fornaro2025filteringST} are recommended, other tracking controllers and estimators can also be used, depending on the required precision and/or implementation limitations.

To evaluate controller performance in terms of power absorption, this study compares Symphony with two state-of-the-art high-level comparators for maximal power absorption: A spectral-based OB controller (SP-Con) \cite{Garcia2019spectral} and the NOB LiTe-Con+ \cite{Garcia2020litecon}\cite{garcia2023broadband}. SP-Con is used as an approximate reference point for maximal performance in constrained, and known, scenarios (alternatively, MPC controllers could also be used). On the other hand, LiTe-Con+ is included as a target for the current-best {NOB} controller performance as studied in~\cite{garcia2020simple}, providing a relevant and complete basis for comparison under different operational conditions. The obtained results show that, without requiring optimisation or complex tuning, Symphony ensures close-to-optimal power absorption across the evaluated peak wave periods, $T_P$, achieving above 90\% of the power absorption obtained with SP-Con for $T_P>5$ s.

The remainder of the paper is structured as follows. In Section~\ref{s:2}, a control-oriented model is presented together with WEC control principles. In Section~\ref{s:3}, the Symphony controller is presented. The results obtained using Symphony are presented in Section~\ref{s:5}, and the main conclusions and findings are summarised in Section~\ref{s:6}.

\section{Control oriented model \& Principles \\ for wave energy converters}\label{s:2}
In this Section, the control-oriented WEC model is presented in Subsection \ref{s:2.1}. Then, in Subsection \ref{s:2.2}, the impedance matching principle for WECs is briefly discussed.

\subsection{WEC modelling fundamentals} \label{s:2.1}
A control-oriented WEC model is obtained, considering a trade-off between abstraction, computational complexity, and precision to represent the system dynamics. Additionally, uncertainties are incorporated for control performance evaluation. Following this reasoning, the control-oriented model used in this paper is obtained by applying Newton's second law, together with linear hydrodynamic assumptions~\cite{falnes2002ocean} to a one-degree-of-freedom (1-DoF) WEC:
\vspace{-0.1cm}
\begin{equation}\label{eq:01_newton}
    m\dot{v} = f_{ex}  + f_u - f_{h}(z) - \Tilde{f}_{r}({v},\dot{v}) + \Delta,
\end{equation}
where $z$, $v$, and $\dot{v}$, represent the WEC displacement, velocity and acceleration, respectively. Also, $m$ is the mass of the device, $f_{ex}$ is the wave excitation force, $f_u$ is the control force, $f_{h}(z) = k_z z$, is the restoring force (with $k_z$ being the restoring coefficient), $\Delta$ represents model uncertainty, and $\Tilde{f}_{r}(v,\dot{v})$, is the radiation force defined \cite{cummins1962impulse} as:
\begin{equation}\label{eq:02_radiation}
    \Tilde{f}_{r}(t) = m_{\infty} \dot{v}(t) + \underbrace{\int_{t_0}^{t} h_r(t-\tau) v(\tau)d\tau}_{f_r(t)},
\end{equation}
where $m_{\infty}$ is the infinite frequency asymptote of the radiation added-mass and $h_{r}(t)$ is the impulse response of the linear convolution operator, which describes the memory effect of the fluid response \cite{perez2008time}. The convolution operator, $f_r(\cdot)$, describes a causal and passive system, and thus can be approximated as a linear, continuous-time, strictly proper, and finite-dimensional system:
\begin{subequations}
\label{eq:radiation_state_space}\begin{align}
    \dot{\mathbf{z}}_r &= \mathbf{F z}_r + \mathbf{G}v, \\
    f_{{r}}   &\approx \mathbf{H z}_r,
\end{align}
\end{subequations}
with $\mathbf{F}\in\Rset^{n_r\times n_r}$, being Hurwitz, $\mathbf{G} \in \Rset^{n_r \times 1}$, and \mbox{$\mathbf{H} \in\Rset^{1 \times n_r}$}. Thus, $\mathbf{z}_r \in\Rset^{n_r}$. For a formal discussion on the properties of $\Sigma_r$, see \cite{perez2008time}. Considering \eqref{eq:01_newton}-\eqref{eq:radiation_state_space}, and assuming that the output is the measured velocity, a complete state space representation of the WEC dynamics is:
\begin{subequations}\label{eq:04_state_space_cummins}
\begin{align}
  \dot{\mathbf{x}}    &= \mathbf{Ax} + \mathbf{B} \left(f_{ex}+f_u + \Delta \right), \\
  \mb{y} = \begin{bmatrix}  x & v \end{bmatrix}\transp   &= \mathbf{Cx} \quad  \left( = \mathcal{G}(f_{ex}+f_u + \Delta) \right),
\end{align}
\end{subequations}
where, with a slight abuse of notation, $\mathcal{G}(\cdot)$ is used to concisely represent the WEC dynamics, $\mathbf{x}^{\intercal}= [z,\, v,\, \mathbf{z}_r^{\intercal}]$, and $\mb{A}$, $\mb{B}$, and $\mb{C}$ are defined as:
\begin{align}\label{eq:SS_description} 
 \!\! \mathbf{A}\! =\! \begin{bmatrix} \mathbf{A}' & \!-\mathbf{B}' \mathbf{H} \\ \mathbf{GC}' & \mathbf{F} \end{bmatrix}, \, \mathbf{B}\! =\! \begin{bmatrix} \mathbf{B}' \\ \mathbf{0} \end{bmatrix}, \mathbf{C}^{\intercal}\! =\! \begin{bmatrix} \mathbf{C}' \\ \mathbf{0} \quad \mathbf{0} \end{bmatrix}, 
\end{align}
with: 
\begin{align}\label{eq:mech_ss_model}
\mathbf{A}' &= \begin{bmatrix}   0 & 1 \\ -\mathcal{M} k_z & 0 \end{bmatrix},\,\mathbf{B}' = \begin{bmatrix}  0 \\ \mathcal{M}  \end{bmatrix}, \, (\mathbf{C}')^{\intercal} = \begin{bmatrix} 1 &  0  \\ 0 & 1 \end{bmatrix},
\end{align}
where $\mathcal{M} = \left(m + m_{\infty}\right)^{-1}$, and the zero vectors $\mathbf{0}\in\Rset^{n_r}$. 

For controller design, without loss of generality, $\Delta = 0$ is assumed. However, $\Delta \neq 0$ is considered for controller evaluation. Specifically, $\Delta$ is used to represent parametric radiation uncertainty.
\subsection{Impedance matching principle} \label{s:2.2}
In wave energy systems, the absorbed mechanical energy is employed for the development of \textit{mechanical energy maximising} control strategies. For system \eqref{eq:04_state_space_cummins}, the integral of converted power over the interval $T \in\Rset_+$ is given by:
\begin{equation}\label{eq:07_cost}
    \mathcal{J}_T = - \int_{T} f_u {v}(\tau) \mathrm{d}\tau.
\end{equation}
In the frequency domain, there exists a closed solution for the problem of maximum power absorption over an infinite-length horizon, i.e., $T\to \infty$. This solution, based on the impedance matching principle \cite{falnes2002ocean}, is obtained by representing the velocity to force response from \eqref{eq:01_newton} in the Fourier domain as:
\begin{equation}\label{eq:08_impedance}
Z(j\omega) = \frac{F_{ex} -F_u}{V} (j\omega),
\end{equation}
where $Z(j\omega)$ is termed the WEC intrinsic impedance, and $F_{ex}$, $F_u$, and $V$, are the Fourier transform pairs of the wave excitation force, control action, and velocity, respectively. Then, the control force required to maximise \eqref{eq:07_cost} may be designed as: 
\begin{equation}\label{eq:09_optimal_control}
    F_u(j\omega) = -V(j\omega) \, Z^{*}(j\omega),
\end{equation}
where $Z^{*}(j\omega)$ represents the complex conjugate of the WEC intrinsic impedance. {Although the solution in \eqref{eq:09_optimal_control} is elegant, this frequency domain approach is not realisable for a panchromatic wave excitation force, since the analytic continuation of \eqref{eq:09_optimal_control} cannot be synthesised as a causal controller \cite{scruggs2010causal}. However, by replacing \eqref{eq:09_optimal_control} in \eqref{eq:08_impedance}, it is possible to obtain the force-to-velocity closed-loop transfer function:
    \begin{equation}\label{eq:imp_match}\begin{split}
         \!\!\!T(j\omega) = \frac{V}{F_{ex}}(j\omega) =  \frac{1}{2 \Re{Z(j\omega)}}\,,
    \end{split}
    \end{equation}
from which the following conditions, employed for the design of controllers for WECs, may be stated:}
    \begin{enumerate}[label = (C\arabic*), ref = {\rm (C\arabic*)}]
        \item \label{Phase_cond} \textbf{Phase condition}: Given that ${1}/{2\Re{Z(j\omega)}}$ is real and even, there is no exchange of reactive energy between {$f_{ex}(t)$} and the controlled WEC, and the optimal WEC velocity is (frequency-wise) proportional to the wave excitation force. 
        \item \label{Amp_cond} \textbf{Amplitude condition}: This dictates the proportionality factor between the excitation force and the controlled velocity. In the time domain, such a factor is provided by the transform pair of $1/{2\Re{Z(j\omega)}}$.
    \end{enumerate}
Utilising conditions \ref{Phase_cond} and \ref{Amp_cond}, various controllers for wave energy systems \cite{ringwood2014energymax}\cite{garcia2020simple}\cite{garcia2023broadband}\cite{Bacelli2020complex}\cite{fusco2012simple} have been designed to approximate, in real-time, the optimal control force from \eqref{eq:09_optimal_control}. Also, suboptimal strategies \cite{babarit2009declutching}\cite{henriques2016latching} have been proposed to achieve \ref{Phase_cond} solely, i.e., to design a controller capable of synchronising the WEC velocity with the wave excitation force. As detailed in the following section, Symphony is designed to approximately satisfy \ref{Phase_cond}. The rationale behind considering \ref{Phase_cond} only can be explained as follows. Simultaneously satisfying conditions (C1) and (C2) results in significant motion amplification of the WECs, leading to reference trajectories that can easily exceed the WEC physical limits \cite{ringwood2023empowering}\cite{Bacelli2013constraints}. Consequently, motion constraints become essential, and, while the amplitude condition \ref{Amp_cond} may not be satisfied, the phase condition \ref{Phase_cond} remains crucial.

\section{Simple phase control (Symphony)}\label{s:3}
\begin{figure}[hb] 
	\centering
	\includegraphics[width = 0.9\columnwidth]{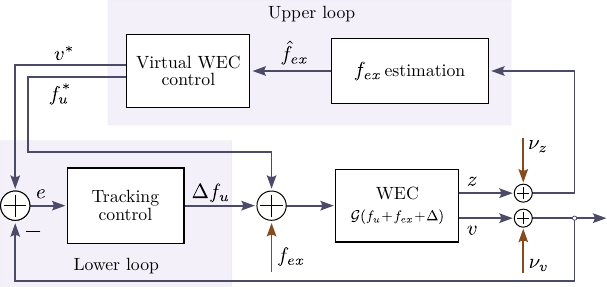}
	\caption{Symphony controller block diagram. Brown arrows denote unknown inputs.}
	\label{fig:control_scheme}
\end{figure}

Symphony operates using a hierarchical structure, illustrated in Figure~\ref{fig:control_scheme}. The upper loop, presented in Subsection~\ref{s:3.1}, provides (i) a feedforward control force and (ii) a velocity reference trajectory that complies with motion constraints. Then, a lower loop, detailed in Subsection~\ref{s:3.2}, tracks the desired velocity reference employing feedback and feedforward control. In addition, a recommended tracking controller and wave excitation force estimator are detailed in Subsection~\ref{s:3.3}.
\subsection{Symphony upper loop}\label{s:3.1}
The upper loop provides reference trajectories and a feedforward control term, which are then used by the lower loop.
\begin{figure}[t]
  \centering
  \includegraphics[width = \columnwidth]{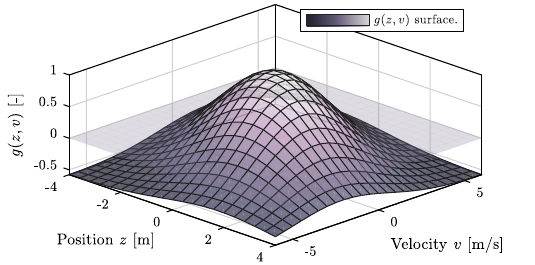}
  \caption{Illustrative $g(z,v)$ using $c = 15$, $z_{M} = 3m$, and $v_{M} = 4m/s$.}
  \label{fig:envelope_x2}
\end{figure}
First, for the design of motion-constrained reference trajectories, assume, without loss of generality, that an estimate of the wave excitation force, $\hat{f}_{ex}(t)$, is available (See Subsection~\ref{s:3.3}). Then, in Symphony, constrained system trajectories are obtained by numerically solving IGDE: 
\begin{equation} \label{eq:opt_vel}
    \dot z = \hat{f}_{ex}(t) \cdot k_{0} \cdot g (z,\dot{z}),
\end{equation}
with $k_0\in \Rset_+$ a user-defined gain and $g(z,\dot{z})$ a Gaussian modulating envelope (GME) defined as: 
\begin{equation}\label{eq:opt_gain}
	g(z,\dot{z}) = \frac{e^{- \left( \frac{z(t)}{z_{M}} \right)^2 - \left( \frac{v(t)}{v_{M}} \right)^2 - \frac{{z(t)} {v(t)}}{c}} -e^{-1}}{1-e^{-1}},
\end{equation} 
with $c, z_{M}, v_{M} \in \Rset_+$ design parameters (observe an illustrative GME in Figure~\ref{fig:envelope_x2}). Solutions of~\eqref{eq:opt_vel} provide motion-constrained reference trajectories, $z\equiv z^*$, and  $\dot{z} \equiv {v}^*$, and, trivially, the feedforward control force $f^*_u$ may be found as the solution to $\dot{z} \equiv {v}^* = \mathcal{G}(\hat{f}_{ex},f^*_u)$, with $\mathcal{G}$ the nominal WEC model in~\eqref{eq:SS_description} (this is further discussed in Section~\ref{s:num_igde}). The following details the essential properties of the Symphony upper loop and design guidelines.
\subsubsection{\underline{Properties of IGDE}} %
Equation~\eqref{eq:opt_vel} possesses appealing characteristics for wave energy control. Specifically, for~\eqref{eq:opt_vel}, the following properties hold:
\begin{enumerate}
    \item The trajectories defined by~\eqref{eq:opt_vel} remain inside a forward invariant (safe) region $\mathcal{D}$, independently of $\hat{f}_{ex}$, meaning that for every $\ms{x}_0\in\mathcal{D}$ ($\ms{x} = [z,\dot z]\transp$), $\ms{x}(t)\in\mathcal{D}$ for $\ms{x}(t_0)= \ms{x}_0$ and all $t > t_0$. $\mathcal{D}$ is given by:
\[\!\!\!\!\!\!\!\!\!\!\mathcal{D} \!=\! \left\{ \ms{x}\in\Rset^2 : |z| \leqslant z_M \wedge \left( \frac{z}{z_M}\right)^2\! +\! \left( \frac{v}{v_M}\right)^2\! +\! \frac{zv}{c} \leqslant 1 \right\}, \]
Set forward invariance of $\mC$ implies that 
\begin{equation} \label{eq:pos_const}
    |z(t)| \leqslant z_{M} \quad \wedge \quad |\dot{z}(t)| \leqslant \frac{2\, v_M\, c }{\sqrt{4c^2 - x_M^2 v_M^2}},
\end{equation}
with $c > \frac{x_M v_M}{2}$, is satisfied $\forall t$. 
\item Since $\mathcal{D}$ is forward invariant, the Symphony upper loop provides trajectories ($\ms{x}$) which comply with motion constraints, independently of $\hat{f}_{ex}$ (which is typically model dependent). This is an important property, particularly considering that, to comply with motion constraints, OB controllers require precise model information and forecasting~\cite{Faedo2021moments}\cite{lin2022fast}.
\end{enumerate}

\subsubsection{\underline{Synthesis of $f_u^*$ and $v^*$}} \label{s:num_igde}
Since~\eqref{eq:opt_vel} is implicit and non-autonomous, to synthesise the velocity reference, $v^*$, and provide a feedforward, $f_u^*$, a numerical approximation obtained by controlling a virtual WEC model is used. Specifically, let $v^*$ be a virtual reference defined as:
\begin{equation} \label{eq:virtual_ref}
v^* \equiv {\hat{f}}_{ex}(t) \cdot k_{0} \cdot g (\Tilde{z},\dot{\Tilde{z}}),
\end{equation}
where $\Tilde{z}$ and $\dot{\Tilde{z}}$ are the position and velocity of a virtual WEC model, i.e., $\dot{\Tilde{z}} = \Tilde{v} = \mathcal{G}(\hat{f}_{ex},f^*_u)$, with $\mathcal{G}$ the nominal WEC model in~\eqref{eq:SS_description}, and $f^*_u$ a virtual control action, designed to steer the error $v^*-\Tilde{v}\to 0$. Hence, using a sufficiently robust tracking controller (such that $v^* \approx \Tilde{v}$ holds), $f^*_u$ is the desired feedforward control action, and $v^*$ is the desired velocity reference.

\begin{figure}[t] 
	\centering
	\includegraphics[width = 0.9\columnwidth]{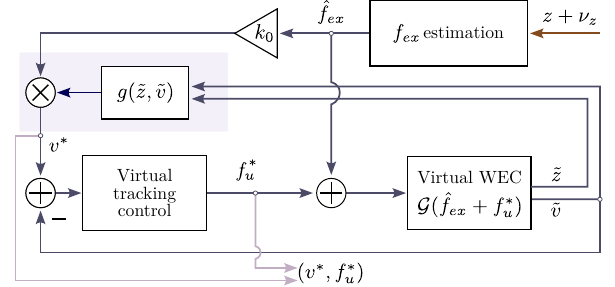}
	\caption{Upper-loop virtual WEC control.}
	\label{fig:control_scheme_ff}
\end{figure}

In this paper, the robust controller employed to track $v^*$~\eqref{eq:virtual_ref}, is the filtering super-twisting (FST), presented in~\cite{fornaro2025filteringST}, and further discussed in Subsection~\ref{s:3.3}. An illustrative scheme representing the closed-loop control of a virtual WEC model, used to solve IGDE, is presented in Figure~\ref{fig:control_scheme_ff}.

\subsubsection{\underline{Tuning parameters of Symphony}}
The parameters to adjust in Symphony are in the GME~\eqref{eq:opt_vel}: $z_M$, $v_M$, $c$, and gain $k_0$. Since $z_M$ and $v_M$ determine hard constraints~\eqref{eq:pos_const}, typically, $z_M$ and $v_M$ are fixed parameters. On the other hand, $c$ and $k_0$ are flexible parameters, which can be determined following simple guidelines. 
\begin{itemize}[leftmargin = 0.4cm]
    \item $k_0$ may be selected using $k_0 > \max \{\frac{1}{2 \Re{Z(j\omega)}}\left|_{\omega\in\Omega} \right. \}$ evaluated across the frequency range of operation, $\Omega$. Alternatively, it is possible to tune $k_0$ adaptively, following, e.g.,~\cite{Fornaro2025symphony}. %
    \item Regarding $c$, the only limitation is $c > \frac{x_M v_M}{2}$, so that $\left( \frac{z}{z_M}\right)^2\! +\! \left( \frac{v}{v_M}\right)^2\! +\! \frac{zv}{c} = 1$ describes an ellipse. $c$ affects Symphony performance in terms of power absorption; however, evaluating the impact of $c$ in the performance of Symphony remains the goal of future research.
\end{itemize}
\begin{Remark}
    In~\eqref{eq:opt_vel}, $\dot{z}\,\propto \,\hat{f}_{ex}$, meaning that the main control objective is to synchronise the WEC velocity with the wave excitation force estimate. The proportionality factor comprises $k_0$, and the GME, $g(z,v)$, included to provide constrained reference trajectories. To guarantee proper operation of the constraint handling mechanism in the virtual WEC model, a robust velocity tracking controller is required, as illustrated in Figure \ref{fig:control_scheme_ff}. In this paper, the FST tracking controller from~\cite{fornaro2025filteringST} is recommended due to: (i) robustness to keep $\Tilde{v}\approx v^*$, and, (ii) since the velocity reference is computed using the estimated (and noisy) excitation force, $\hat{f}_{ex}$, the filtering structure in FST considerably reduces the effects of noise. 
\end{Remark}
\vspace{-0.2cm}
\begin{Remark}
	Although the procedure to synthesise $f_u^*$ requires a WEC model, the velocity reference, $v^*$, obtained solving IGDE~\eqref{eq:opt_vel}, complies with motion constraints independently of the WEC model, and hence, independently of model errors, since the only model dependency in~\eqref{eq:opt_vel}, introduced via $\hat{f}_{ex}$, does not affect the set forward invariance of~\eqref{eq:opt_vel}.
\end{Remark}

\vspace{-0.3cm}
\subsection{Symphony lower loop}\label{s:3.2}
In Symphony, the lower loop tracks the safe trajectory references given by~\eqref{eq:opt_vel}, by combining feedforward and feedback terms. The main advantage behind the inclusion of a feedforward control force is as follows. With Lipschitz bounded model uncertainty, $\Delta$, typically found in WECs, including $f^*_u(t)$ reduces the control effort applied by the tracking controller. This permits reducing the tracking control gain, subsequently enhancing noise resilience. For clarity sake, the recommended tracking controller for Symphony is presented in Subsection~\ref{s:3.3}. 

It is worth noting that, to design the lower loop, different state-of-the-art topologies for control of wave energy systems can be used. For instance, it is possible to resort to a pure feedforward structure, as in~\cite{Faedo2021moments}\cite{garcia2023broadband}, a closed-loop feedback velocity tracking structure, as in, e.g.,~\cite{fusco2014hierarchical}, or a combination of feedforward and feedback~\cite{Merigaud2018horizon}\cite{Faedo2024sm2}. If the uncertainty is small, using only feedforward control action $f_u^*$ may be sufficient to comply with motion constraints and power performance standards. However, arguably, as studied in~\cite{Merigaud2018horizon}\cite{Faedo2024sm2}, using only feedforward control action has limited applicability in realistic scenarios, and including a feedback loop is crucial to comply with motion constraints. 

The Symphony lower loop is instrumental in achieving practical compliance with motion constraints. Since the reference trajectories, $\ms{x}^* = [z^*, v^*]\transp$, comply with $\ms{x}^*\in\mathcal{D}$, with $\mathcal{D}$ the safe set for~(11), a security margin, $\Delta_{z,v}$, introduced to account for tracking error, $e$, which depends on the employed tracking controller, is sufficient to comply with motion constraints. As a particular example, with tracking error $e$, the WEC trajectories are: 
\( v = v^* + e, \) and 
\( z = \int (v^* + e)\, \mathrm{d}\tau.\)
Hence, using $z_M = z_{\mathrm{max}} - \Delta_z $, with $z_{\mathrm{max}}$ the maximum admissible displacement, and $\Delta_z$ a security margin introduced to account for tracking error, is sufficient to comply with motion constraints.
\begin{Remark}\label{rmk:01}
    In WEC controllers, which incorporate a velocity tracking loop, the performance, in terms of power absorption, is determined by the velocity reference, $v^*$; hence, the performance, in terms of power absorption, is (in)sensitive to WEC model errors only if $v^*$ is (in)sensitive to WEC model errors. This occurs even in the presence of a feedforward component, since $f_u^*$ is treated as a disturbance and rejected by the SM tracking controller.
\end{Remark}
Note that, in Symphony, $v^*$ is given by~\eqref{eq:opt_vel}, and the only dependency with the WEC model is introduced via the wave excitation force estimate, $\hat{f}_{ex}$. Although a sensitivity study is beyond the scope of the present paper, as shown in Section~\ref{s:5}, Symphony performance, in terms of power absorption, remains close to optimal even in the presence of radiation model errors.

\subsection{Tracking controller and wave excitation force estimation}\label{s:3.3}
Wave excitation force estimation and velocity tracking control are essential for Symphony operation. Although multiple alternatives are available, due to their robustness to disturbances and measurement noise, SM-based alternatives are recommended. Specifically, two SM-based filtering algorithms for $f_{ex}$ estimation and velocity tracking are presented in the following.
\subsubsection{Wave excitation force estimation}
To robustly estimate the wave excitation force, a homogeneous SM-based unknown input observer (HSMO), presented in~\cite{fornaro2024homogeneous}, is used.

In essence, HSMO is a mimic of the WEC model~\eqref{eq:SS_description}, which is built assuming that a measurement of the position, $z$, is available. Also, it is assumed that the measured position contains Lebesgue measurable noise, $\nu_z$, with a bounded second-order integral, i.e., $\iint \nu_z< \varepsilon_z \in \Rset_+$. Then, HSMO is formulated as:
\begin{subequations}\label{eq:11}\begin{align}
    \dot{w}_{f1}         &= w_2 + k_{h1} \lceil w_{f1} \rfloor^{4/5} ,                                           \label{eq:11.1}  \\
    \dot{w}_{f2}     &= \hat{z} - (z+\nu_z) + k_{h2} \lceil w_{f1} \rfloor^{3/5},                                       \label{eq:11.2} \\
    \dot{\hat{z}}     &= \hat{v} + k_{h3} \lceil w_{f1} \rfloor^{2/5},                                           \label{eq:11.3}  \\
    \dot{\hat{v}}     &= \hat{f}_{ex}\mathcal{M} - (\hat{f}_r + \hat{f}_h + f_u)\mathcal{M} + k_{h4} \lceil w_{f1} \rfloor^{1/5}, \label{eq:11.4}  \\
    \dot{\hat{f}}_{ex} &= k_{h5}/\mathcal{M} \lceil w_{f1} \rfloor^{0},                                                     \label{eq:11.5}  \\
    \dot{\hat{\mb{x}}}_r   &= \mb{F}\hat{\mb{z}}_r + \mb{G}\hat{v} ,                              \label{eq:11.6}  \\
    \hat{\Tilde{f}}_r &= \mb{H}\hat{\mb{z}}_r,                                                   \label{eq:11.7}  \\
    \hat{f}_h         &= k_z\hat{z} ,                                                       \label{eq:11.8}
\end{align}\end{subequations}
where the notation $\lceil \cdot \rfloor^{j} = |\cdot|^{j} \text{sign}(\cdot)$ is employed. \eqref{eq:11.1}--\eqref{eq:11.2} constitute a $2$-order SM filter, designed to integrate the error $e_{z}= \hat{z} - (z + \nu_z)$, and reduce the influence of $\nu_z$. Also, $k_{hi}$, with $i=1,2,...,5$, are gains parametrised in terms of a Lipschitz bound-based constant, $L_h\in\Rset_+$. Specifically, following~\cite{Levant2020}:
\begin{align}\label{eq:gains_3}\begin{array}{lll}
    \!k_{h1} = 5\cdot L_h^{1/5},  &   \!k_{h2} = 10.03 \cdot L_h^{2/5}, & \!k_{h3} = 9.3\cdot L_h^{3/5}, \, \\ 
    \!k_{h4} = 4.57 \cdot L_h^{4/5}, &    \!k_{h5} = 1.1\cdot L_h. &  \nonumber
\end{array}
\end{align}
Under a simple boundedness assumption, $L_h$ exists, and it can be found via in-silico evaluations; we refer the interested reader to~\cite{fornaro2024homogeneous} for further details on gain selection. Recall that Symphony does not require HSMO specifically, and other estimators (some discussed in~\cite{pena2019critical} and references therein) can be used.

\subsubsection{Tracking control}
To robustly track velocity profiles, \pedro{a wide variety of tracking controllers can be implemented}. In this paper, due to its robustness, FST SM control is employed~\cite{fornaro2025filteringST}. FST, used for velocity tracking in both Symphony upper and lower loops\pedro{, requires a relative degree ($\rho$) of the sliding variable ($v-v^*$), with respect to the control action, of 1}\footnote{\pedro{Different} S\pedro{M} \pedro{algorithms have} already been experimentally validated as tools for WEC control~\cite{Faedo2024sm2}\pedro{\cite{Mosquera2023gst}}, \pedro{which supports the practical applicability of assuming $\rho=1$ for velocity tracking}. If $\rho\neq1$ (due to, for instance, \pedro{unmodelled} actuator dynamics), \textit{nested} SM \pedro{approaches~\cite{Fornaro2026host}} can be used.}.

\pedro{In} FST, chattering is reduced by using a higher-order version of the ST controller, which includes (i) a filtering structure to reject measurement noise, $\nu_v$, satisfying $\iint \nu_v< \varepsilon_v \in \Rset_+$ as detailed in~\cite{fornaro2025filteringST}\cite{Levant2020}\cite{Fornaro2026host}, and (ii) the discontinuous action in the second time derivative of the control force. Specifically, the FST control law is: 
\begin{subequations}\label{eq:controlSM_noise}\begin{align}
          \dot{w}_1    &= w_2   + k_{c1} \lceil w_1 \rfloor^{4/5}              \\
          \dot{w}_2    &= (v + \nu_v)-v^* + k_{c2} \lceil w_1 \rfloor^{3/5}    \\
           \Delta f_u  &= u_2 + k_{c3} \lceil w_1 \rfloor^{2/5}, \\
        \dot{u}_2      &= u_3 + k_{c4} \lceil w_1 \rfloor^{1/5},             \\
        \dot{u}_3      &= k_{c5} \lceil w_1 \rfloor^{0},
    \end{align} \end{subequations}
where $\Delta f_u$ is the FST control action, the parameters $k_{ci}$, $i=1,2,...,5$, are gains parametrised in terms of a Lipschitz bound-based gain $L_c\in\Rset_+$, and, as in the HSMO, $w_{1,2}$ are auxiliary variables, included to filter the measured velocity and reject Lebesgue measurable noise $\nu_v$. This latter noise-rejection feature proves essential for preserving the tracking performance of FST by considerably reducing the influence of measurement noise, $\nu_v$. Following \cite{Levant2020}, an appropriate selection of the gains $k_{ci}$, $i=1,2,...,5$, is:
\begin{align}\label{eq:gains_2}\begin{array}{ll}
    k_{c1} = 5\cdot L_c^{1/5}, & k_{c2} = 10.03 \cdot L_c^{2/5}, \\
    k_{c3} = 9.3\cdot L_c^{3/5} /\mathcal{M}, & k_{c4} = 4.57 \cdot L_c^{4/5} /\mathcal{M}, \\
    k_{c5} = 1.1\cdot L_c /\mathcal{M},  &
\end{array}
\end{align}
selected with $L_c$ obtained empirically via numerical evaluation. Employing~\eqref{eq:gains_2}, \pedro{and assuming fast actuator and sensor dynamics with respect to the WEC dynamics,} after some finite time, $T$, the system remains, robustly, on \pedro{a small neighbourhood of} the sliding manifold $\forall t>T$~\cite{Orlov2009discontinuous}\cite{Levant2016robustness}. Following~\cite{Levant2020}, in the presence of measurement noise $\nu_v$ satisfying $\iint \nu_v< \varepsilon_v \in \Rset_+$ and a maximal sampling time $\tau$, the tracking error, $e$, around the origin, for $t > T$, satisfies
    \[| e | = e_M < \mu_v L_c \kappa^{5},\quad \kappa = \max[ (\varepsilon_v / L_c )^{\frac{1}{5}}, \tau ], \]
for some $\mu_v>0$, and $\forall t \geqslant t_0$~\cite{Levant2020}. \pedro{In practice, the bound on $e$ also depends on the contribution of sensor, actuator, and other unmodelled dynamics, as studied in~\cite{MendozaAvila2023parasitic}}. Although conservative, $e_M$ could be used to inform the Symphony upper loop, to reduce $z_M$ and $v_M$ in order to comply with motion constraints. 


\section{Symphony controller numerical evaluation}\label{s:5}
In this section, in-silico studies are conducted to evaluate Symphony performance in terms of power absorption. To that end, simulation assumptions and requirements are articulated in Subsection~\ref{s:5.1}. Then, results without model uncertainty are presented in Subsection~\ref{s:5.2}, and results with parametric model uncertainty and measurement noise are analysed in Subsection~\ref{s:5.3}. Importantly, recall that the obtained results are preliminary and limited to a 1-DoF WEC.

\subsection{Preliminaries}\label{s:5.1}
In this section, the assumptions required to assess Symphony performance in terms of power absorption are presented and briefly discussed.
\subsubsection{Nominal WEC model}
\begin{figure}[h]
  \centering
  \includegraphics[width = \columnwidth]{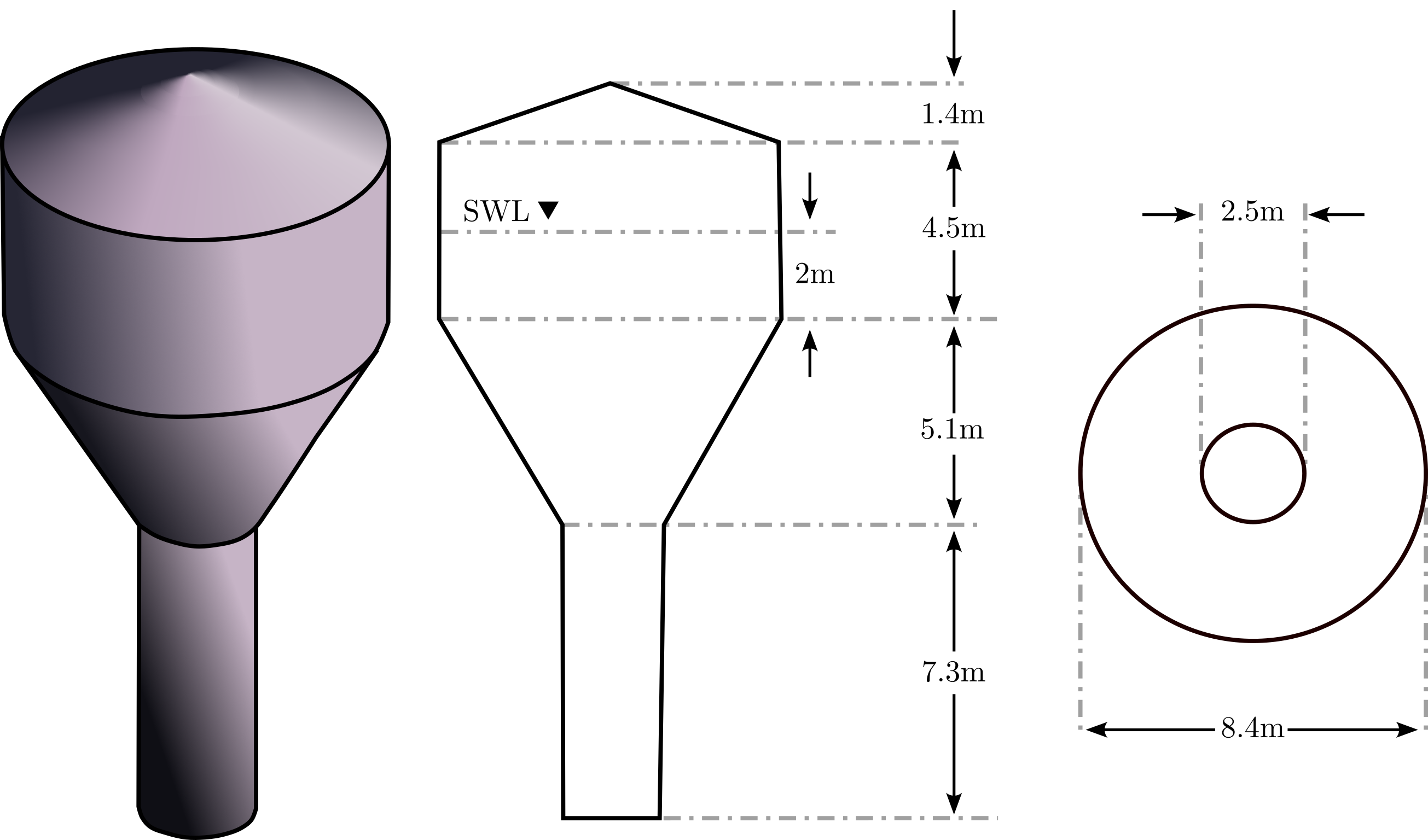}
  \caption{Schematic illustration of the point absorber WEC device. SWL stands for `still water level'.}
  \label{fig:device_schematic}
\end{figure}

The Symphony controller is evaluated on a CorePower-like WEC device~\cite{Todalshaug2016tank}, illustrated in Figure \ref{fig:device_schematic} together with the system dimensions. By assuming linear hydrodynamics, and considering a 1-DoF device constrained to heave motion (movement in the $z$ axis only), the WEC dynamics are given by \eqref{eq:04_state_space_cummins}, where $\mathcal{M}$ and $k_x$ are presented in Table \ref{tab:01}, the zero vectors are $\mathbf{0}\in\Rset^{7}$, and $\mb{G}$, $\mb{H}$ and $\mb{F}$ are given by: 
\begin{subequations}\label{eq:radiation_matrix}\begin{align}
\!\mb{F} &= \begin{bmatrix}
-7&	\!-24&	\!-47&	\!-57&	\!-43&	\!-18&	\!-3.4 \\
1&	  0&	0&	   0&	0& 	0&	0            \\                                                                                                         
0&	  1&    0&	   0&	0&	0&	0            \\
0&	  0&	1&	   0&	0&	0&	0            \\
0&	  0&	0&	   1&	0&	0&	0            \\
0&	  0&	0&	   0&	1&	0&	0            \\
0&	  0&	0&	   0&	0&	1&	0            \\
\end{bmatrix}\!, \\
\!\mb{G} &= \begin{bmatrix} 1.46 \cdot 10^{5} & 0 & 0 & 0 & 0 & 0 & 0 \end{bmatrix}\transp\!, \label{eq:24b_rad}\\
\!\mb{H} &= \begin{bmatrix} 0.21 &  \!1.1 &  \!3.8 & \!4.6  & \!3 & \!0.64 & \!0.014 \end{bmatrix} \!\cdot\!- 10^{-5}\!.
\end{align}\end{subequations}

\begin{table}[h]\small
\caption{Nominal system model and SM control parameters}
\label{tab:01}
\begin{center}
\begin{tabular}{|| m{2.2cm} | m{2.2cm} | m{2.2cm} ||}
\hline
\multicolumn{3}{||c||}{\scriptsize \textbf{WEC parameters }} \\
\hline
\hline
\centering \scriptsize$\mathcal{M} = 6.8\times 10^{-6}$ & \centering \scriptsize $k_x = 5.57 \times 10^{5}$ & \multicolumn{1}{c||}{\scriptsize $n_r = 7$}   \\
\hline
\hline
\centering \scriptsize{\textbf{Parameter}} & \multicolumn{2}{ c||}{\scriptsize \textbf{SM controller and estimator}}      \\
\hline
\hline
\scriptsize$L_h$     &       \multicolumn{2}{c||}{\scriptsize     $6$ $[m/s^3]$ }                      \\
\scriptsize$L_c$     &       \multicolumn{2}{c||}{\scriptsize    $ 100$  $[m/s^3]$ }                 \\
\hline
\end{tabular}
\end{center}
\end{table}\normalsize

\subsubsection{Sea realisations}
In this paper, without loss of generality, the JONSWAP spectrum \cite{hasselmann1973measurements} is employed. To obtain statistically representative results, for each sea state, 30 realisations of 2000s are obtained by white noise filtering. The sea states are obtained considering peak periods $T_p \in [5,5.5,6,..., 12.0]s$, a significant wave height $H_s = 2.6 m$, and a steepness parameter $\gamma =3.3$. %
\subsubsection{Symphony controller configurations}
For the WEC under study, assume that the position constraint is: 
\begin{align}
        |z(t)|   &\leqslant 2m = z_M, \label{eq:const_max_pos}
\end{align}
and no assumptions are made regarding $v_M$. Therefore, for constrained scenarios, the GME, $g(z,v)$, is designed using $z_M = 2m$, and $c = 15$. Then, four tuning alternatives for Symphony are analysed: 
\begin{itemize}
    \item \textbf{S1:} $v_m = 2.3$ m/s, $k_0 = 3.5 \cdot 10^{-5}$ m/$($s$\cdot$N$)$.
    \item \textbf{S2:} $v_m = 2.3$ m/s, $k_0 = 4.2 \cdot 10^{-5}$ m/$($s$\cdot$N$)$.
    \item \textbf{S3:} $v_m = 5.2$ m/s, $k_0 = 3.5 \cdot 10^{-5}$ m/$($s$\cdot$N$)$.
    \item \textbf{S4:} $v_m = 5.2$ m/s, $k_0 = 4.2 \cdot 10^{-5}$ m/$($s$\cdot$N$)$.
\end{itemize}
The differences between \textbf{S1}--\textbf{S4} are in the GME parameter $v_M$, and constant gain $k_0$. As shown in the results analysis, the value of $k_0$ contributes marginally to the performance of \textbf{S1}--\textbf{S4}, while $v_M$ is a critical parameter. The employed SM tracking controller and estimator parameters are presented in Table~\ref{tab:01}. 

\subsubsection{Benchmark controllers for comparison}
In order to critically assess the results obtained with the Symphony controller, two benchmark controllers are employed:
\begin{itemize}
    \item First, an OB formulation, the SP-Con controller~\cite{Garcia2019spectral} is used to obtain a target reference for the maximum power absorption achievable in constrained. Note that, without considering model uncertainty, any OB control, such as MPC~\cite{lin2022fast} or moment-based control~\cite{Faedo2021moments}, can be used to obtain an approximate optimal constrained solution. %
    \item  Second, to compare the results of the Symphony controller with a state-of-the-art {NOB} controller, the LiTe-Con+ is used \cite{Garcia2020litecon}\cite{garcia2023broadband}. As mentioned in~\cite{garcia2020simple}, LiTe-Con+ obtains the higher levels of mechanical power absorption among a wide class of NOB controllers.%
\end{itemize}

\subsubsection{Radiation uncertainty and measurement noise}
To evaluate the robustness of Symphony to uncertainty in the radiation model, evaluations in constrained scenarios are obtained considering a parametric error in the radiation model. Specifically, it is assumed that the \textit{model} used for $f_{ex}$ estimation has an error $\Delta$, such that $\hat{\mb{G}} = \mb{G} (1 + \Delta)$, where $\mb{G}$ is in~\eqref{eq:24b_rad} and $\Delta \in \{ -0.3,-0.25, ... , 0.25, 0.3\}$. 

Additionally, sensor measurement noise is assumed for both position and velocity. Specifically, following~\cite{pena2019critical}, $\nu_z \sim \mathcal{N}(0, 9\cdot 10^{-5})$  and $\nu_v \sim \mathcal{N}(0, 2.5\cdot 10^{-4})$ are selected as representative cases of realistic sensor noise levels.

\begin{figure}[t]
  \centering
  \includegraphics[width = \columnwidth]{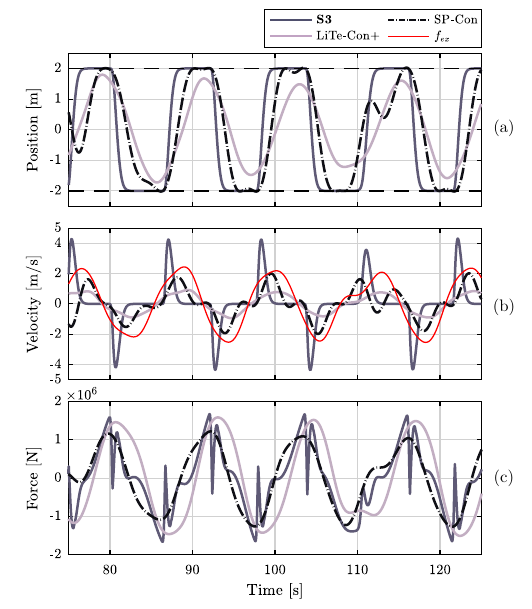}
  \caption{Time evolution of the WEC variables in the constrained case. Sea realisation with $T_p=12s$.}
  \label{fig:PC_05}
\end{figure}
\subsection{Symphony performance in idealised constrained scenarios}\label{s:5.2}
In this section, time-domain solutions are presented, followed by an evaluation of the Symphony performance in terms of average power absorption, with no model uncertainty ($\Delta = 0$) or measurement noise ($\nu_z = \nu_v = 0$).

\subsubsection{Time domain analysis}\label{s:5.3.1}
In this section, the time evolution of the controller variables is analysed, considering a realisation with $T_p=12$. Specifically, the time domain solutions of LiTe-Con+ and SP-Con are compared with the Symphony configurations \textbf{S1} and \textbf{S3}.

\begin{figure}[t]
	\centering
	\includegraphics[width = \columnwidth]{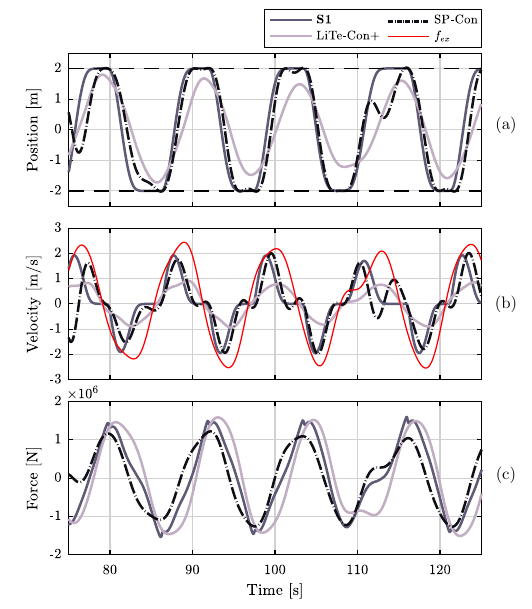}
	\caption{Time evolution of the WEC variables in the constrained case. Sea realisation with $T_p=12s$.}
	\label{fig:PC_052}
\end{figure}

\begin{figure}[h]
	\centering
	\includegraphics[width = \columnwidth]{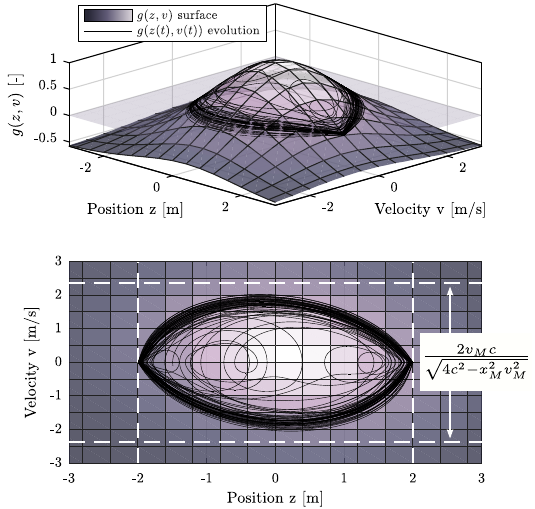}
	\caption{Time evolution of the GME gain $g(z(t),v(t))$ on the modulation manifold $g(z,v)$. Sea realisation with $T_p=12s$.}
	\label{fig:PC_surf}
\end{figure}
Observe, first, in Figure \ref{fig:PC_05}, time domain solutions using \textbf{S3} ($v_M=5.2$ m/s). It can be seen, in Figure \ref{fig:PC_05} (top), that each controller satisfies constraint \eqref{eq:const_max_pos}, with SP-Con and \textbf{S3} solutions approaching the constraint limit with less conservativism than LiTe-Con+. Additionally, note that \textbf{S3} solutions resemble a latching strategy: The WEC velocity is forced to zero when the WEC position reaches the constraint, $z_M$. However, due to a large $v_M$ in \textbf{S3}, the system is rapidly released when the system goes away from the position constraint limit, producing large velocity peaks (see Figure \ref{fig:PC_05} (middle)), which are correlated with high-frequency peaks in the control variable (Figure \ref{fig:PC_05} (bottom), whereas both SP-Con and LiTe-Con+ control forces possess smooth time variations.

{However, by analysing \textbf{S1} time domain variables, it can be noted that designing $g(z,v)$ with $v_M= 2.3m/s$} reduces both the maximum speed limits and the rapid variations in the control force (see Figure \ref{fig:PC_052}). This can be appreciated in Figure \ref{fig:PC_052} (middle), where, for the same time interval and sea realisation as in Figure \ref{fig:PC_05}, the maximum velocity is reduced from approximately $4m/s$ to $2m/s$. As a result, the control force is smooth, without high-frequency peaks introduced to latch the WEC. Hence, the $v_M$ parameter plays a fundamental role in shaping the $f_u$ time evolution, with lower velocity limits resulting in reduced control force variations.

{To evaluate the satisfaction of position and velocity constraints, the evolution of $g(z,v)$, for \textbf{S1}, is presented in Figure \ref{fig:PC_surf}. In the position-velocity phase plane, it can be noted that position and velocity constraints are satisfied. Also, in Figure \ref{fig:PC_surf}, it can be appreciated how $g(z,v)$ guarantees satisfaction of the constraints for both the WEC position and velocity, by forcing $g(z,v)\to0$, as the operational space of the controlled variables increases, i.e., as $z\to z_M$ and $v\to v_M$.}
\begin{figure}[t]
	\centering
	\includegraphics[width = \columnwidth]{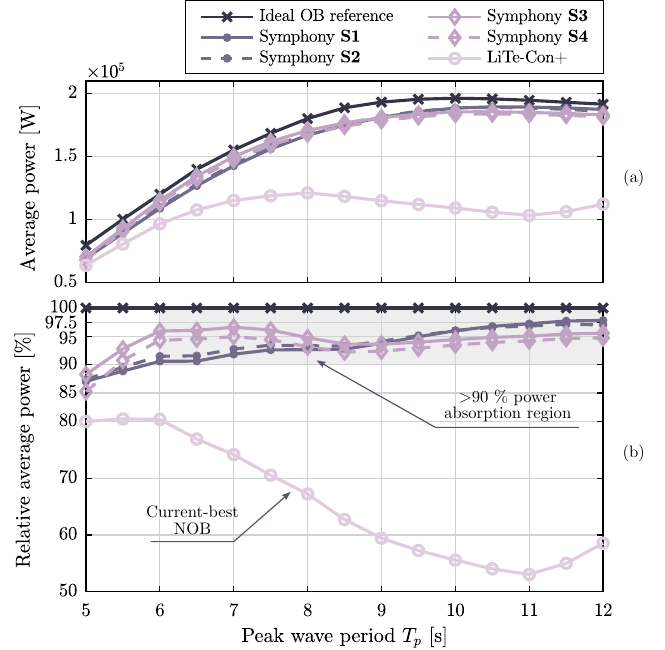}
	\caption{Average power extraction in the constrained case. Comparison between \textbf{S1}, \textbf{S3}, and the ideal OB reference SP-Con.}
	\label{fig:RPC_083}
\end{figure}
\subsubsection{Power absorption performance}
In this section, \textbf{S1}--\textbf{S4} performance is presented and compared in terms of power absorption. The Symphony versions are compared with the idealised OB reference, obtained using SP-Con (Recall that SP-Con is not a real-time control strategy~\cite{Merigaud2018horizon}), and LiTe-Con+.

In Figure \ref{fig:RPC_083}, it can be observed and compared the average power absorption of the evaluated controllers. Notably, the averaged power absorption obtained with Symphony is close to optimal for \textbf{S1}--\textbf{S2}, above 87.5\% for the evaluated sea states, while LiTe-Con+ provides, approximately, between 60-80\% on average, but up to 60\% of average power absorption for sea states with larger $T_p$. The difference in power absorption between LiTe-Con+ and Symphony may be explained as follows: While both LiTe-Con+ and Symphony handle position constraints, the Symphony constraint handling mechanism enables a larger state space exploration. Then, by exploiting a larger dynamical range, Symphony possesses a performance close to the theoretical optimum and significantly enhances the average power extraction obtained with LiTe-Con+.

It is particularly interesting to compare \textbf{S1}-\textbf{S2} performance with \textbf{S3}-\textbf{S4} performance. While the $k_0$ value has an approximate 2\% contribution for power output, $v_M$ selection is critical: Sea states with smaller $T_p$ require larger $v_M$ to maximise power absorption. While beyond the scope of the present paper, an adaptive $v_M$ selection depending on sea state characteristics may be implemented. Regardless, interestingly, the performance of every Symphony configuration is still close to optimal, specifically, above 90$\%$ for most of the evaluated sea states, and reaching 98\% for \textbf{S1} and $T_p\approx 12$~s.

Finally, Symphony is compared with SP-Con. While the SP formulation requires a precise model of the system and information over a large time window to provide an optimal control force, for its part, Symphony is {NOB} and, to comply with motion constraints, independent of the system parameters. Additionally, it is well-known that, to satisfy the hard position constraints, SP-Con may become computationally intractable for real-time operation; conversely, Symphony, in addition to its close-to-optimal performance, is capable of robustly satisfying position constraint~\eqref{eq:const_max_pos} without the necessity of adjusting any control parameter. 

\subsection{Symphony performance in realistic constrained scenarios}\label{s:5.3}
Although close-to-optimal for the evaluated sea states, in Section~\ref{s:5.2}, Symphony performance, in terms of power absorption, is assessed assuming perfect knowledge of the wave excitation force and measured signals. Hence, this section now evaluates Symphony separately considering (i) measurement noise for position and velocity, and (ii) uncertainty in the radiation model (for conciseness, only \textbf{S1} and \textbf{S3} are evaluated).

\subsubsection{Power performance with measurement noise} %
\begin{figure}[t]
	\centering
	\includegraphics[width = \columnwidth]{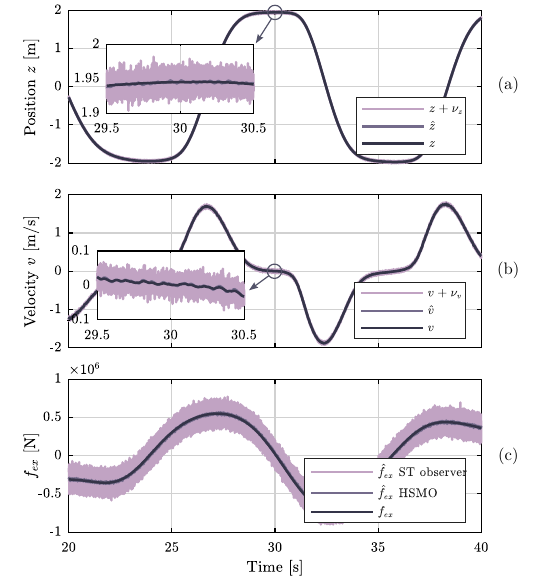}
	\caption{Variables estimated using HOSM with noisy measurements. (a) Position. (b) Velocity. (c) Excitation force.}
	\label{fig:noisy_vars}
\end{figure}

Consider the case where both measured position and velocity are contaminated with measurement noise, i.e., $z_m = z + \nu_z$ and $v_m = v + \nu_v$, with $\nu_z \sim \mathcal{N}(0, 9\cdot 10^{-5})$ and $\nu_z \sim \mathcal{N}(0, 2.5\cdot 10^{-4})$, selected as realistic cases for sensor measurement noise, as detailed in~\cite{pena2019critical}. Note that $\nu_{z,v}$ are signals of infinite filtering order, hence complying with the assumptions required for the design of the wave excitation force estimator~\eqref{eq:11} and control~~\eqref{eq:controlSM_noise}. First, observe noisy signals in Figure~\ref{fig:noisy_vars}. The estimated variables $\hat{z}$ and $\hat{v}$ are obtained using HSMO, and it can be observed, in Figures~\ref{fig:noisy_vars}.a and~\ref{fig:noisy_vars}.b, that the effect of noise is ameliorated. Importantly, the estimated excitation force $\hat{f}_{ex}$ is presented in Figure~\ref{fig:noisy_vars}.c.c, where a super-twisting SM observer~\cite{mosquera2024sliding} is compared with HSMO. Here, it is possible to appreciate that the HSMO filtering structure possesses a positive impact on $f_{ex}$ estimation.

\begin{figure}[t]
	\centering
	\includegraphics[width = \columnwidth]{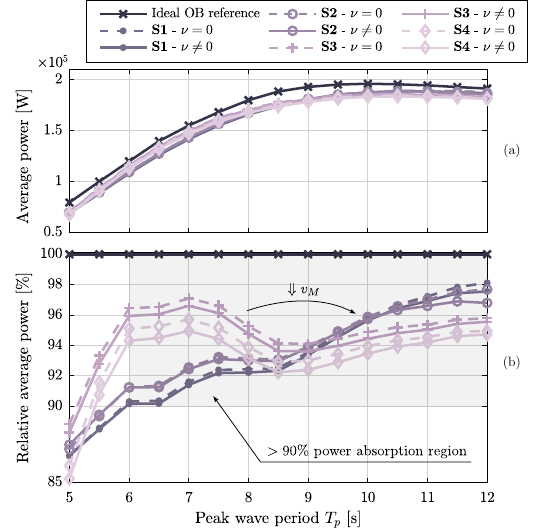}
	\caption{Average power absorption in constrained scenarios. Comparison between \textbf{S1}, \textbf{S3}, and the ideal OB reference SP-Con.}
	\label{fig:noisy_power}
\end{figure}

In terms of power absorption, results are presented in Figure~\ref{fig:noisy_power}. Notably, Symphony performance remains close to optimal despite measurement noise, which diminishes power absorption by (approximately) 1\%. In addition, recall that measurement noise also affects the performance of OB controllers, with an impact on maximum power output.

\subsubsection{Power performance with $\Delta \neq 0$} %
\begin{figure}[t]
	\centering
	\includegraphics[width = \columnwidth]{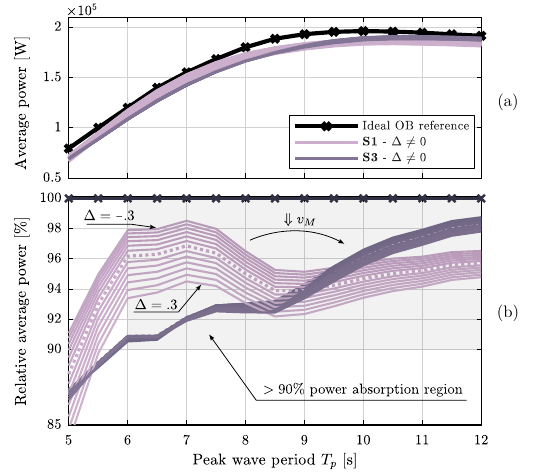}
	\caption{Average power absorption in constrained scenarios. Comparison between \textbf{S1}, \textbf{S3}, and the ideal OB reference SP-Con.}
	\label{fig:rad_unc}
\end{figure}

The performance, in terms of power absorption, of controllers \textbf{S1} and \textbf{S3}, is separately computed for the cases $\Delta \in \{ -0.3,-0.25, ... , 0.25, 0.3\}$. The obtained results are presented and superimposed in Figure~\ref{fig:rad_unc}. Notably, Symphony remains close-to-optimal: Less than 4\% variation in power absorption is achieved for \textbf{S3}, which indicates that Symphony is \textit{largely} insensitive to radiation errors. Interestingly, with a variation in average power absorption less than 1\%, \textbf{S1} shows more resilience to radiation errors than \textbf{S3}.

\begin{figure}[t]
	\centering
	\includegraphics[width = \columnwidth]{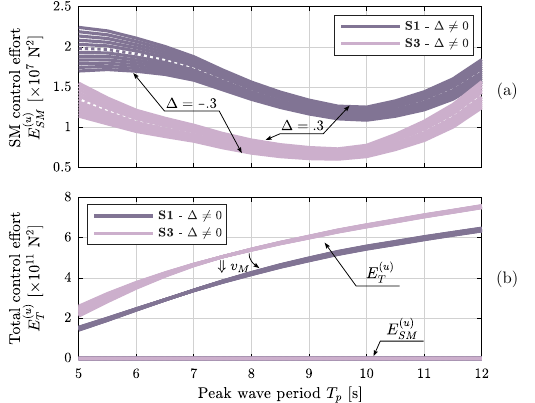}
	\caption{Control effort contributions with radiation errors.}
	\label{fig:rad_control_effort}
\end{figure}

In addition, it is also interesting to evaluate the contribution of feedback control action, $\Delta f_u$, particularly in comparison with total control action $\Delta f_u+f_u^*$. To that end, the control efforts are computed as time-averaged squared control forces:
		\[ E^{(u)}_{S\pedro{M}} = \frac{1}{t} \cdot \int^t_0 \Delta f_{u}^2 \,\mathrm{d}\tau,  \quad E^{(u)}_{T} = \frac{1}{t} \cdot \int^t_0 ( \Delta f_{u} + f_u^*)^2 \,\mathrm{d}\tau. \]
The obtained results, presented in Figure~\ref{fig:rad_control_effort}, are averaged across multiple realisations and computed \pedro{for} varying radiation errors. Notably, the SM effort is four orders of magnitude smaller than the feedforward control term, evenly across realisations with varying radiation errors, evidencing that the SM tracking controller provides a small contribution to track the desired velocity reference, even in the presence of a radiation error reaching $\pm30\%$.

It is also interesting to analyse the differences between \textbf{S1} and \textbf{S3}. Since \textbf{S3} is tuned with a larger $v_M$, a larger control effort is required to track the IGDE trajectories. Also, since $E^{(u)}$ averages the control effort, this metric does not reflect that, as presented in Figures~\ref{fig:PC_05} and~\ref{fig:PC_052}, using \textbf{S3}, the control force presents high-frequency peaks introduced to latch the WEC when the maximum position constraints are reached. Hence, in summary, $v_M$ is a critical parameter, since its value determines Symphony power performance and the tracking control effort.

\begin{figure}[t]
	\centering
	\includegraphics[width = \columnwidth]{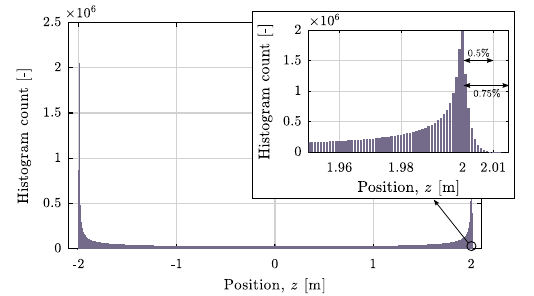}
	\caption{Histogram count across 13500 realisations considering radiation errors and measurement noise. Adding a security margin $\Delta_z = 0.015 m$ would suffice to guarantee constraint satisfaction.}
	\label{fig:rad_bins}
\end{figure}

To evaluate constraint satisfaction, a histogram that considers $13500$ realisations (including measurement noise and radiation errors) is presented in Figure~\ref{fig:rad_bins}. In accordance with the theoretical set forward invariance of IGDE~\eqref{eq:opt_vel}, the reference velocity does not exceed constraints. However, due to a small tracking error, modest constraint violations are expected. To provide a quantitative evaluation, the average exceedance per sample, $E_T^{(\%)}$, is computed as: 
\[  E_T^{(\%)} = \frac{100}{N} \cdot \!\!\! \sum_{i:\,|z_i| > z_M} \!\!\!\!(|z_i| - z_M)\, n_i,\]
where $z_i$ are histogram bin centres, $n_i$ are bin counts, and $N = \sum_i n_i$. Considering $13500$ realisations, $E_T^{(\%)} = 0.025\%$ is obtained, with a maximum value of $z = 2.015$ m. Hence, designing the GME using $z_M = 1.985 m$, would suffice to guarantee constraint satisfaction under the evaluated scenarios.
Empirically,  a security margin, $\Delta_z$, may be selected depending on the robustness of the tracking controller employed. Also, in Figure~\ref{fig:rad_bins}, it is interesting to note a concentration of samples close to $z_M$, illustrating how the GME mechanism exploits the operational space of the WEC within the constraints.
\section{Conclusions} \label{s:6}

In this paper, Symphony, a novel {NOB} controller for wave energy systems, is presented. The proposed controller operates by tracking a modulated velocity reference proportional to the excitation force. Specifically, the main innovation of Symphony is in obtaining constrained trajectory references via solving an IGDE. In essence, IGDE solutions lie within a forward-invariant subset of the state space; hence, the resulting trajectories guarantee phase synchronisation between the excitation force estimate and the WEC velocity, and motion constraints satisfaction. In addition, unlike in OB formulations, IGDE trajectories are independent of the WEC model and forecast is not required to comply with motion constraints.

The obtained results exemplify the efficacy of the proposal: In terms of power absorption, Symphony is the first {NOB} controller to achieve close-to-optimal performance, in constrained scenarios, with minimum recalibration. Specifically, Symphony achieves, in most of the evaluated scenarios, more than 90\% of the power absorption obtained by a spectral (optimisation-based) control solution. Symphony is the first {NOB} controller capable of providing a velocity reference that complies with motion constraints, independently of WEC model errors and measurement noise. Moreover, although a WEC model is required to implement Symphony, its performance, in terms of power absorption, is largely insensitive to radiation errors. In summary, Symphony represents a new competitive alternative for real-time WEC control, not only for its simplicity and robustness, but also for its close-to-optimal constrained performance.

Some critical areas for future development include conducting thorough experimental evaluations, designing a hard-force constraint mechanism, extending the Symphony controller principle of operation for \pedro{decentralised} WEC array control, evaluating its use for multiple-DoF WECs\pedro{, and developing an electric-energy-maximising Symphony version}.

\bibliographystyle{IEEEtran}
\bibliography{bC1}
\end{document}